\documentstyle[aps,twocolumn]{revtex}

\input{epsf}

\def\be{\begin{eqnarray}}
\def\ee{\end{eqnarray}}

\def\g{\gamma}

\def\o{\omega}

\begin{document}
\title{Nonlinear Hall effect in a time-periodic electric
\ field and related phenomena.}
\author{A.A.~Ovchinnikov}
\address{{\it Max Planck Institute for Physics of Complex Systems, Dresden;\\
Joint Institute of Chemical Physics, RAS, Moscow.}}

\wideabs{
\maketitle

\begin{abstract}

A motion of neutral and charged particles located nearby a metal surface under
joint action of time-periodic electric field $E(t)$ directed normally to
the surface and permanent magnetic field $H$ directed along the surface has
been considered. It has been shown that due to a nonlinearity of the system
there exists a directed transport of the particles perpendicularly to both
magnetic and electric fields with velocity which is proportional to $E^2H$.
The velocity has been evaluated for charged (ions, electrons etc.) and
neutral (atoms, molecules, nano-size clusters etc.) particles adsorbed on a
metal surface. Corresponding surface electric current has been found as a
function of a frequency of the electric field and material parameters. It
has been noted that the similar phenomena appear in a bulk of electrically
inhomogeneous media.
\end{abstract}
}


The directed transport of particles under action of a time-periodic force is
now well established phenomenon in various non-linear systems. The details
of this transport has been studied for one-dimensional motion of particle in
the space-periodic potential $U(x)$ and a time periodic force having general
form $F(t)=F_{0}\cos {\omega t}+F_{1}\cos {(2\omega t+\theta )}$. (See
review \cite{1} and recent works \cite{2,3,4,5}.) Main physical reason for
such directed motion of the particles is due to the so called latent
asymmetry of driving force $F(t)$. Though the time average of $F(t)$ is zero
the time average $\overline{F^{3}(t)}$ doesn't vanish and depends on $\theta
$. Nonlinearity of the system leads to mixing of harmonics such a way that
average velocity is not zero and directed along $\overline{F^{3}(t)}$.

The aim of this paper is to show that the similar time- asymmetry arises in
a permanent magnetic field even in the case of purely harmonic external
force $F(t)$. The phenomenon is closely related with a conventional Hall
effect in conducting media. In difference with the latter the DC electric
current is arising on interface under a joint action of the harmonic
electric field directed normally to the surface and a permanent magnetic
field directed along the surface. The effect is of the second order with
respect to the amplitude of the electric field. For this reason we'll call
it the nonlinear Hall (NHL) effect.

Firstly we calculate the transverse velocity of a single charged particle
located on a metallic surface in the AC electric field and longitudinal
permanent magnetic field. Then, in Sections 2,3 we consider the similar
phenomenon in a surface electron gas of metals and semiconductors. In
Section 4 we evaluate the transport velocity of neutral particles in above
conditions. Finally we give a phenomenological description of the NLH effect
in inhomogeneous conducting media (Section 5).

\section{Directed motion of single charged particle.}

In order to demonstrate the physical mechanism of DC current let us consider
a motion of a single ion with a mass $M$ on metallic surface. A complete
description of an interaction between the ion and metallic surface is a
quite complicated problem. In framework of a density functional approach it
has been discussed in a recent monograph by Liebsch \cite{6}. There are a
few solidly established facts we are going to use in our paper. The
interaction between the ion and the metallic surface consists mainly of two
parts. The first part is an attractive Coulomb force between the ion and an
image charge having a potential $V_{im}(z)$
\begin{equation}
V_{im}(z)=-Q^{2}/(2z\epsilon )  \label{1}
\end{equation}
where $z$ is a distance between the ion and an edge of the metal surface and
$\epsilon $ is a static dielectric constant. We gave here the simplest
expression for such force at large $z$ neglecting the dynamics of the metal
electrons. More rigorous consideration includes an integration over the
complex surface polarisability, which modifies the $z$- dependence of $%
V_{im}(z)$. For the sake of simplicity we will use (1) with some
dimensionless parameter $c_{1}$
\begin{equation}
V_{im}(z)=-c_{1}Q^{2}/z  \label{2}
\end{equation}
The second part of the interaction is connected with a short-range repulsive
exchange potential $V_{ex}(z)$ which we choose in exponential form
\begin{equation}
V_{ex}(z)=c_{2}a_{1}\exp \{-(z-z_{e})/a_{1}\}  \label{3}
\end{equation}
were $z_{e}$ is an equilibrium distance between the phy\-sisorbed ion and
metal surface, $a_{1}$ is a radius of exchange forces on surface. The $c_{2}$
is chosen from a equilibrium condition
\begin{equation}
V(z)=V_{im}(z)+V_{ex}(z)  \label{4}
\end{equation}
\begin{equation}
\partial V/\partial z|_{z=z_{e}}=0  \label{5}
\end{equation}
\begin{equation}
c_{2}=c_{1}Q^{2}/z_{e}^{2}  \label{6}
\end{equation}
The motion of the physisorbed ion with a velocity $\nu $ leads to electronic
excitations in the metal and therefore to a friction force for such motion.

The friction force may be written as \cite{6}
\begin{equation}
{\bf f}=-M(\eta _{\parallel }\nu _{\parallel }+\eta _{\perp }\nu _{\perp })
\label{7}
\end{equation}
Here ${\bf \nu }$ is the particle velocity and $\eta _{\parallel }$, $\eta
_{\perp }$ are the friction coefficients for the motion parallel and normal
to the surface. If the charge is located outside the electronic density of
metal, it can be shown that $\eta _{\parallel }=\eta _{\perp }/2$ \cite{6}.
At large ion surface distances $z$ the friction parameter $\nu _{\perp }$
depends on $z$ as
\begin{equation}
\nu _{\perp }={\frac{{Q^{2}}}{M}}{\frac{{\omega _{F}}}{{k_{F}\omega _{p}^{2}l%
}}}{\frac{1}{{z^{3}}}}  \label{8}
\end{equation}
where $\hbar \omega _{F},k_{F},l$ and $\omega _{p}$ are a Fermi energy, a
Fermi momentum, a mean free path of electrons and a plasmon frequency
respectively.

For rough estimation it is convenient to use more simple formula
\begin{equation}
\nu _{\perp }={\frac{{Q^{2}}}{{M\sigma _{0}z^{3}}}}  \label{9}
\end{equation}
Here $\sigma _{0}$ is a bulk conductivity of a metal. The latter gives a
correct order of magnitude for $\nu _{\parallel }$ and $\nu _{\perp }$.

Having in mind all these facts one can readily write the equations of motion
of the ion under a joint action of the AC electric field $E_{0}\cos {\omega t%
}$ directed along $z$ (normally to the surface) and a permanent magnetic
field directed parallel to the surface along y-axis, $H_{y}=H$.

\begin{equation}
M{\ddot{z}}=-M\nu (z){\dot{z}}+{\frac{1}{c}}Q{\dot{x}}H-{\frac{\partial V}{%
\partial z}}+\xi _{z}(t)+QE_{0}(t)  \label{10}
\end{equation}
\begin{equation}
M{\ddot{x}}=-Mv_{\parallel }(z){\dot{x}}+{\frac{Q{\dot{z}}H}{c}}+\xi _{x}(t)
\label{11}
\end{equation}
Here $c$ is a light velocity. We included also into these equations the
random forces $\xi _{x(z)}(t)$ which reproduce the thermal fluctuations of
the ion velocity. As usual, they have the white noise correlations of the
following type

\begin{eqnarray}
&<&\xi _{x}(t_{1})\xi _{x}(t_{2})>=2\nu _{\parallel }(z)k_{B}T\delta
(t_{1}-t_{2})  \nonumber \\
&<&\xi _{z}(t_{1})\xi _{z}(t_{2})>=2\nu _{\perp }(z)k_{B}T\delta
(t_{1}-t_{2})  \label{12}
\end{eqnarray}
where $k_{B}$ is a Boltzmann constant and $T$ is a temperature of the system.

In spite of the comparative complexity of these equations they can be
analyzed analytically in the limit of the very small electric and magnetic
fields. At small amplitudes $E_{0}$, $H$ of external fields a displacement
of the ion $\delta z=z-z_{e}$ from its equilibrium position is quite small
and the first equation (10) could be chosen in a linear form
\begin{equation}
M{\ddot{z}}=-2M{\dot{z}}\nu _{0}-k_{e}\delta z+QE_{0}(t)  \label{13}
\end{equation}
where $k_{e}$ is a force constant of a vibration of ion along the z-axis
\begin{equation}
k_{e}={\frac{\partial ^{2}V}{\partial z^{2}}}\Bigr|_{z=z_{e}}  \label{14}
\end{equation}
and
\begin{equation}
\nu _{0}=2\nu _{\perp }(z_{e})  \label{15}
\end{equation}
We neglected in (10) the thermal fluctuations ($T=0$) and the Lorentz force
term. This is correct if we are interested in a linear contribution with
respect to magnetic field into the velocity of the transport. However we
have to keep this term in (11).

The nonlinearity of Eq.(11) is connected with damping term, which in
accepted approximation could be written as
\begin{equation}
-M\nu _{0}(1-\gamma \delta z){\dot{x}}  \label{16}
\end{equation}
Here $\gamma $ is a derivative of the damping function $\nu $ taken at
equilibrium distance $z_{e}$. Then, the Eq.(11) took the following form
\begin{equation}
M{\ddot{x}}=-M\nu _{0}(1-\gamma \delta z){\dot{x}}+{\frac{QH}{c}}{\dot{z}}
\label{17}
\end{equation}
The periodic solution of (13) is
\begin{equation}
\delta z(t)=Re{\Bigl\{{\frac{E_{0}(Q/M)e^{i\omega t}}{{\omega
_{e}^{2}-\omega ^{2}+i\nu _{0}2\omega }}}\Bigr\}}  \label{18}
\end{equation}
Here $\omega _{e}$ is an eigen-frequency of the physisorbed ion, $\omega
_{e}^{2}=k_{e}/M$. Then the periodic solution of the second linear equation
(17) could be written as the following

\begin{equation}
\begin{array}{ll}
{\dot{x}}=&\int_{0}^{\infty }d\xi \lbrack QH/cM]\exp [-\xi \nu _{0}]\delta {%
\dot{z}}(t-\xi )\times \\
&~~~\exp {\{-\gamma \nu _{0}\int_{t-\xi }^{t}\delta z(\tau
)d\tau \}}\\
\end{array}
\label{19}
\end{equation}
We are interested in the transverse velocity ${\dot{x}}$ averaged over the
period of the vibration in the external electric field $E_{0}(t)$. Making
use the necessary averaging we have a final expression
\begin{equation}
{\bar{v}}(\omega )={\frac{Q^{3}E_{0}^{2}H\gamma }{2cM^{3}}}{\frac{\omega ^{2}%
}{{\omega ^{2}+{\bf \nu }_{0}^{2}}}}{\frac{1}{{(\omega _{e}^{2}-\omega
^{2})^{2}+4\omega ^{2}{\bf \nu }_{0}^{2}}}}  \label{20}
\end{equation}
The function ${\overline{v}}(\omega )$ has a resonance character and reaches
its maximal value at $\omega =\omega _{e}$. Since $\nu _{0}\sim M^{-1}$ and
$\omega _{e}^{2}\sim M^{-1}$ the ${\overline{v}}(\omega _{e})$ depends quite
weakly on the mass of ion. As a rule, $\nu _{0}\ll \omega _{e}$ (damping is
comparatively small) and
\begin{equation}
\overline{v}(\omega _{e})={\frac{Q^{3}E_{0}^{2}H\gamma }{2cM^{3}4\omega
_{e}^{2}\nu _{0}^{2}}}
\label{21}
\end{equation}
The behavior of $\overline{v}(\omega )$ as function of $\omega $ is given in
Fig.1 for three different values of mass. On the same figure we gave the
results of numerical solutions of Eqs.(10,11).

Let us estimate the possible velocities $\overline{v}$ at reasonable
experimental parameters
$$
\begin{array}{ll}
M\sim 10^{-24}g,~~&\gamma \sim 10^{8}cm^{-1},\\
\omega_{e}\sim 10^{12}sec^{-1},~~ &\nu_{0}\sim 0.01\omega_{e}\\
\end{array}
$$
and $Q$ is an electron charge. Then
\[
\overline{v}(\omega _{e})\sim 10^{-3}E_{0}^{2}H,
\]
where $E_{0}$ and $H$ have to be taken in emu system. For example, if $H\sim
10^{3}Gauss$ and $E_{0}\sim 100V/cm$, $\overline{v}(\omega _{e})\sim 1$
cm/sec.

In order to clarify the physical source of the directed transport of the
particle we give a qualitative analysis of the equations(13) and (17). First
of all we neglect the dependence of the damping coefficient $\nu (z)$ on $z$%
, $\gamma =0$. Then, system of equations (13) and (17) became a linear one.
The solution of this system can be written as following

\begin{eqnarray}
\delta z &=&a_{0}(\omega )\cos {(\omega t+\varphi _{z}(\omega ))}  \nonumber
\\
x(t) &=&a_{0}(\omega )b_{0}(\omega )\cos {(\omega t+\varphi _{z}(\omega
)+\varphi _{x}(\omega ))}  \label{22}
\end{eqnarray}
where $a_{0}(\omega )$ and $\varphi _{z}(\omega )$ are modulus and argument
of a complex function
\[
{\frac{E_{0}Q}{M}}{\frac{1}{{\omega _{e}^{2}-\omega ^{2}+i2\nu _{0}\omega }}%
}
\]
and $b_{0}(\omega )$ and $\varphi _{x}(\omega )$ are modulus and argument of
a complex function ${%
{\displaystyle{\omega _{L} \over {\nu _{0}+i\omega }}}%
}$.

A closed trajectory of the particle in the plane $(\delta z,x)$ is an
ellipse obeyed the following equation (Fig.2)
\begin{equation}
{\frac{(x-\delta zb_{0}\cos {\varphi _{x}})^{2}}{{a_{0}^{2}b_{0}^{2}\sin ^{2}%
{\varphi _{x}}}}}+{\frac{\delta z^{2}}{a_{0}^{2}}}=1  \label{23}
\end{equation}
If either $\omega _{L}/\nu _{0}$ or $\omega /\nu _{0}$ trends to zero the
ellipse turns into a line. An angle $\psi $ between axes $z$ and a largest
axis of the ellipse is proportional to $b_{0}(\omega )$. Points $A$ and $%
A^{\prime }$ on the Fig.2 are the points of return of the trajectory in the $%
x$-direction. At points $C$ and $C^{\prime }$ the trajectiry cross the line $%
\delta z=0$, and dashed lines $AB$ and $A^{\prime }B^{\prime }$ are parallel
to this line. In fact, the ellipse (23) is an attractor of the system
(13),(17).

Consider a small deformation of the trajectory taking into account the
dependence of $\eta (z)$ on $z$. According to (16)the damping coefficient $%
\eta (z)$ is higher at $\delta z<0$ ($\gamma >0$) and lower at $\delta z>0$.
After switching on this dependence the trajectory of the particle between
points $C$ and $B$ will be shifted a little bit to the right side. But this
shift will be compensated completely by the shift to the left side on $%
AC^{\prime }$ part of the trajectory. The same is true for $AC$ and $%
C^{\prime }B^{\prime }$ parts of the trajectory. Non-zero shift originates
from $AB$ and $A^{\prime }B^{\prime }$ parts. Indeed the velocity on the $AB$
part is positive and the trajectory will be shifted to the right side if we
diminish the damping coefficient of the particle. The velocity on $A^{\prime
}B^{\prime }$ is negative and if we increase the damping the trajectory will
be shifted also to the right side. All in all it leads to the shift of whole
trajectory to the right which is equivalent to the transport of particle in
this direction. The qualitative consideration given above does not depends
on details of the attractor and holds for any non-linear dynamics of the
system at.

Averaging the Eq.(17) over period of time we arrive to the following
equality for the shift in $x$-direction
\begin{equation}
{\overline{\delta x}}=\gamma \int \delta z{\dot{x}}dt  \label{24}
\end{equation}
The integral in (24) is nothing but an area of the ellipse $S$ and we have a
very simple formula
\begin{equation}
{\overline{\delta x}}=\gamma S  \label{25}
\end{equation}
Eq.(25) could be used directly for the calculation of ${\overline{v}}(\omega
)$ and of course it leads to the same result as (20).

Two additional remarks to the main expression (20). If the external AC
electric field is a sum of two electric fields with two different amplitudes
and frequencies $\omega _{1}$ and $\omega _{2}$ the effect is additive
\[
{\overline{v}}={\overline{v}}_{1}(\omega _{1})+{\overline{v}}_{2}(\omega
_{2})
\]
where $v_{1(2)}(\omega _{1(2)})$ are average velocities (20) with $%
E_{0}=E_{1(2)}$ and $\omega =\omega _{1(2)}$, correspondingly. Moreover, any
non-thermal electric noise in system leads to a directed transport of
charged particles. Corresponding average velocity could be obtained by
integration of (20) over a frequency distribution of the electric noise.

Now, if we have two weakly coupled particles with opposite charges and
different masses $M_{1}$ and $M_{2}$ the average velocity of their mass
centre can be defined as
\[
{\overline{v}}(\omega )={\frac{{M_{1}\overline{{v_{1}}}(\omega )-M_{2}%
\overline{{v_{2}}}(\omega )}}{{M_{1}+M_{2}}}}
\]
leading to some neutral current in system. More realistic situation with
strongly coupled charged particles will be considered in Section 4.

\section{The NLH effect on metal surface.}

The similar phenomenon takes play inside a metal on the boundary separating
two media having different conductivities. In this section we consider the
metal-insulator or metal-vacuum boundary as simplest example of such
interface. Consider dynamics of electronic gas in surface layer of metal
under an influence of external AC electric field $E_0(t)=E_0\cos{\omega t}$
directed perpendicularly to surface of the metal (along $z$ axis) and in the
magnetic field $H$ directed along the surface, axis $y$. At this geometry a
density of electronic gas, $\rho (z,t)$, is a function of normal coordinate $%
z$ and a time $t$. There are two velocity components of the electronic gas:
the normal component $v(z,t)$ and the tangential component $u(z,t)$ both of
which are functions of only $z$ and $t$. The electronic gas vibrates
perpendicularly to the surface. Due to the nonlinearity of this vibrations a
normal velocity $v(z,t)$ and density of electron gas $\rho (z,t)$ are
periodic functions of time with frequency $\omega$ having all overtones:
\[
v=v_1(z)\cos{(\omega t+\varphi_1)}+v_2(z)\cos{(2\omega t+\varphi_2)}\ldots
\]
\[
\rho=\rho_0+\rho_1(z)\cos{(\omega t+\varphi_3)}+\rho_2(z)\cos{(2\omega
t+\varphi_4)}\ldots
\]
The above amplitudes and phases of harmonics as functions of $z$ are a
subject of calculations. The Lorentz forces induced by magnetic field lead
to tangential vibration of electronic gas with amplitude which is
proportional to permanent magnetic field. Corresponding tangential velocity $%
u(z,t)$ has the same form as $v(z,t)$ with its own amplitudes and phases.
The density of tangential current is $j(z,t)=u(z,t)\rho(z,t)$. The time
averaging of $j(z,t)$ over period gives the DC component of the current.

Since the width of a surface layer and all other characteristic lengths
greater than atomic distances one can readily use hydrodynamics approach
(see \cite{6,7}) for description of system. In planar geometry all variables
are the functions of only one coordinate $z$ and a periodic functions of $t$
with frequency $\omega $. The main equations look as following \cite{6,7}
\begin{eqnarray}
&&\rho \Biggl({\frac{{\partial v}}{{\partial t}}}+v{\frac{{\partial v}}{{%
\partial z}}}\Biggr)=-{\frac{{\partial p}}{{\partial \rho }}}{\frac{{%
\partial \rho }}{{\partial z}}}+{\frac{{2\pi e}}{{m}}}Q(z)  \nonumber \\
&-&{\frac{{e}}{{m\mu }}}\rho v-{\frac{e}{m}}E_{0}(t)\rho -{\frac{{He}}{{mc}}}%
\rho u+\eta {\frac{{\partial ^{2}v}}{{\partial z^{2}}}}  \label{26}
\end{eqnarray}
\begin{equation}
\rho \Biggl({\frac{{\partial u}}{{\partial t}}}+v{\frac{{\partial u}}{{%
\partial z}}}\Biggr)={\frac{{He}}{{mc}}}\rho v-{\frac{{e}}{{m\mu }}}\rho
u+\eta {\frac{{\partial ^{2}u}}{{\partial z^{2}}}}  \label{27}
\end{equation}
\begin{equation}
{\frac{{\partial \rho }}{{\partial t}}}=-{\frac{{\partial }}{{\partial z}}}%
(v\rho )  \label{28}
\end{equation}
\begin{equation}
Q(z,t)=2q(z)-q(\infty )  \label{29}
\end{equation}
Here $p$ is a pressure of electronic gas in metal, $\mu $ is a mobility of
electron in metal, c is the light velocity, m is effective mass of
electrons, $\eta $ is a viscosity coefficient of an electron fluid. A
quantity
\begin{equation}
q(z,t)=e\int_{0}^{z}(\rho (z^{\prime },t)-\rho _{0})dz^{\prime }  \label{30}
\end{equation}
is a surface density of charge (together with positive charge of a lattice)
accumulated in a layer between $z=0$ and $z$. The boundary conditions for
this system are quite simple. Both velocities $v,u$ are equal to zero at the
edge of metal ($z=0$).

We are looking for a periodic solution of the equations (22-25). The surface
electric current directed along the surface and averaged over the period of
time is
\begin{equation}
I_{H}=\int_{0}^{\infty }e{\overline{\rho (z,t)u(z,t)}}dz  \label{31}
\end{equation}
The bar in Eq.(31) means the time averaging.

The equation (28) can be rewritten via the $q(z,t)$ as
\begin{equation}
{\frac{dq}{dt}}=-ev(z,t)\rho (z,t)  \label{32}
\end{equation}
From this equation we see immediately that the time averaged normal
component of the current
\[
{\overline{j}_{z}}(z)=0
\]
Then, using the Eq.(27) we can get the following formula for time averaged
tangential component of the current
\begin{equation}
\omega _{\mu }{\overline{j}_{x}(z)}=-e{\frac{\partial }{\partial z}}[{%
\overline{\rho (z,t)u(z,t)v(z)}}]+e{\overline{\eta {\frac{\partial ^{2}u}{%
\partial z^{2}}}}}  \label{33}
\end{equation}
The full surface electric current $I_{H}$ can be obtained by integration
over $z$ and looks as following
\begin{equation}
I_{H}=-{\frac{e\rho _{0}}{\omega _{\mu }}}{\overline{u_{\infty }v_{\infty }}}%
-{\frac{e}{m\omega _{\mu }}}{\frac{\partial \eta }{\partial \rho }}%
\int_{0}^{\infty }{\overline{{\frac{\partial \rho }{{\partial z}}}{\frac{%
\partial u}{{\partial z}}}}}dz  \label{34}
\end{equation}
where $v_{\infty }(t)$ and $u_{\infty }(t)$ are the normal and tangential
velocities for large $z$. The constant
\[
\kappa =m/{%
{\displaystyle{\partial \eta  \over \partial \rho }}%
}
\]
has a dimension of a kinematic viscosity $cm^{2}/sec$. At small electric and
magnetic fields first non-vanishing contribution to the $I_{H}$ can be
obtained if we know $q(z,t)$, $u(z,t)$ and $v(z,t)$ in a linear
approximation.

Taking the limit $z\rightarrow \infty $ we come to three equations for
asymptotic values $q_{\infty }=q(\infty ,t)$, $u_{\infty }=u(\infty ,t)$, $%
v_{\infty }=v(\infty ,t)$
\begin{equation}
{\frac{{d}}{dt}}q_{\infty }=-\rho _{0}v_{\imath }  \label{35}
\end{equation}
\begin{equation}
{\frac{{d}}{dt}}v_{\infty }=-v_{\imath }\omega _{\mu }+{\frac{e}{m}}%
E_{0}(t)-2\beta q_{\imath }(t)  \label{36}
\end{equation}
\begin{equation}
{\frac{{d}}{dt}}u_{\infty }=v_{\imath }\omega _{L}-u_{\imath }\omega _{\mu }
\label{37}
\end{equation}
Here $\omega _{\mu }=e/m\mu $, $\beta ={2\pi e^{2}/m}$ and $\omega
_{L}=eH/mc $ is a Larmor frequency. Periodic solutions of these linear
equations give the boundary conditions at $z\rightarrow \infty $. And $%
q(0,t)=0$ by definition. Since one needs only the periodic solution of
(26-29) the initial conditions are arbitrary. The system of equations
(26-29) has been analyzed both by perturbation theory with respect to
external electric field and numerically using standard NAG routines.

The periodic solution of linear equations (35-37) can be written in
following form

\begin{eqnarray}
v_{\infty }(t) &=&c_{e}Re[a(\omega )e^{i\omega t}]  \nonumber \\
q_{\infty }(t) &=&[\epsilon _{0}c_{e}/\omega _{\mu }]\rho _{0}Re[b(\omega
)e^{i\omega t}]  \label{38} \\
\eta _{\infty }(t) &=&\epsilon _{0}hc_{e}Re[c(\omega )e^{i\omega t}]
\nonumber
\end{eqnarray}
Here we introduced a dimensionless functions
\begin{eqnarray}
a(\omega ) &=&{\frac{{\omega _{\mu }\omega }\epsilon _{0}}{{(\omega
_{s}^{2}-\omega ^{2})+i\omega \omega _{{_{\mu }}}}}}  \nonumber \\
b(\omega ) &=&{\frac{{i\epsilon _{0}\omega _{\mu }^{2}}}{{(\omega
_{s}^{2}-\omega ^{2})+i\omega \omega _{{_{\mu }}}}}}  \label{39} \\
c(\omega ) &=&a(\omega ){\frac{{\omega }_{{L}}}{{\omega _{\mu }+i\omega }}}%
\text{{,}}  \nonumber
\end{eqnarray}
where $\epsilon _{0}$ and $h$ are the dimensionless electric and magnetic
fields
\[
\epsilon _{0}={\frac{eE_{0}}{{\omega _{\mu }c_{e}m}}},~~~h={\frac{eH}{{%
mc\omega _{\mu }}}},~~~\omega _{s}^{2}={\frac{2\pi e^{2}\rho _{0}}{m}}
\]
Using (26-29) as boundary conditions at an infinite $z$ we arrive to the
following solutions of the Eqs.(26-29)

\begin{eqnarray}
v(z,t) &=&c_{e}Re[a(\omega )e^{i\omega t}(1-e^{-\xi k})]  \nonumber \\
q(z,t) &=&\rho _{0}Re[b(\omega )e^{i\omega t}(1-e^{-\xi k})]  \label{40} \\
u(z,t) &=&c_{e}Re[c(\omega )e^{i\omega t}(1-e^{-\xi k})]  \nonumber
\end{eqnarray}

Here $\xi $ is a dimensionless coordinate, $z=c_{e}\xi /\omega _{\mu }$ ,
and complex $k=k_{1}+ik_{2}$ is defined from the equation
\begin{equation}
(k_{1}+ik_{2})^{2}=[2\omega _{s}^{2}-\omega ^{2}+i\omega \omega _{\mu
}]/\omega _{\mu }^{2},~~k_{2}>0,  \label{41}
\end{equation}
We are choosing a root which has $k_{2}>0$. Omitting intermediate
calculations we give a final expression for $I_{H}$ in following form
\begin{equation}
I_{H}=-{\frac{{ec_{e}^{2}\rho _{0}}}{{2\omega _{\mu }}}}h\epsilon {_{0}}%
^{2}\gamma _{1}(\omega )+{\frac{\kappa \rho _{0}e}{2}}\gamma _{2}(\omega );
\label{42}
\end{equation}
$\gamma _{1}$ and $\gamma _{2}$ are dimensionless functions of a frequency $%
\omega $ of the external field
\[
\begin{array}{ll}
\gamma _{1}(\omega )= & a^{\ast }(\omega )c(\omega )+c^{\ast }(\omega
)a(\omega ); \\
\gamma _{2}(\omega )= & [k^{2}k^{\ast }c^{\ast }(\omega )b(\omega )+h.c.][{%
k+k^{\ast }}]^{-1}
\end{array}
\]
Both of them have resonance character and reach their peaks at $\omega
=\omega _{s}$ as it shown on Fig.3. Maximal value of them are $\gamma
_{1}(\omega _{s})=2\omega _{\mu }^{2}/\omega _{s}^{2}$ and $\gamma
_{2}(\omega _{s})=1$. For metals at low temperature as a rule $\omega _{\mu
}^{2}/\omega _{s}^{2}\ll 1$. At very high frequency $\omega \gg \omega _{s}$
both $\gamma _{1}(\omega )$ and $\gamma _{2}(\omega )$ decrease as $1/\omega
^{4}$. At low frequency $\omega \ll \omega _{\mu}$ they behave as $\omega
^{2}$: $\gamma _{1}(\omega )\sim \omega _{\mu }^{2}\omega ^{2}/\omega
_{s}^{4}$ and $\gamma _{2}(\omega )=\omega ^{2}/\omega _{s}^{2}$.

Thus, at low temperature main contribution to the surface current $I_{H}$
comes from the second term of (38), depending on the viscosity of electronic
gas. According to \cite{7a} the viscosity of electronic Fermi liquid in
metal at low temperature could be estimated as
\begin{equation}
\eta \cong m\rho _{0}c_{e}d(\epsilon _{F}/T)^{2}  \label{43}
\end{equation}
where $d$ is a lattice constant and $\epsilon _{F}$ is a Fermi energy of a
metal.

Thus, at low temperature the surface current $I_{H}$ can be expressed as
\[
I_{H}=-{\frac{\omega ^{2}e\rho _{0}dc_{e}}{4\omega _{s}^{2}}}\Biggl({\frac{%
\epsilon _{F}}{k_{B}T}}\Biggr)^{2}\Biggl({\frac{\mu E}{c_{e}}}\Biggr)^{2}%
\Biggl({\frac{\mu M}{c}}\Biggr)
\]
It is easy to see that the temperature dependence of the effect is
determined by the factor $\sigma ^{3}(T)/T^{2}$ ($\sigma (T)$ is a
conductivity of the metal).

To demonstrate the value of the effect let us take material parameters of
pure copper at low temperature, say, T=30~K:
\[
d=10^{-8}~cm,~~~~c_{e}=10^{8}~{\frac{cm}{sec}},~~~~\rho
_{0}=10^{22}~cm^{-3}.
\]
The conductivity behave as $T^{-5}$:
\[
\sigma (T)=\sigma _{R}(T_{R}/T)^{5}
\]
where $\sigma _{R}=10^{6}~om^{-1}$ is a room temperature conductivity and $%
T_{R}=300~K$.

Finally, for the pure cooper at $T=300~K$ we have
\[
I_{H}=-0.1\nu ^{2}E_{0}^{2}H
\]
where $\nu $ is expressed in MHz, $E_{0}$ in a V/cm, $H$ in a $Tesla$, $I_{H}$
is in a $\mu A/cm$. At $E_{0}=10~V/cm$, $H=0.01~T$, $\nu =10~MHz$ one has $%
I_{H}=\sim 1.0~\mu A/cm$.

\section{NLH current in an injection layer of a semiconductor.}

In the previous Section we have considered the dynamics of\ electron gas
inside a metal \ on the boundary metal-semiconductor. If a work function for
transition of electrons from metal side to the semiconductor is not too
high electrons of the metal can be injected into the semiconductor creating
an injection layer of electrons inside of the semiconductor. The injection
layer can give a contribution to the NLH current on the interface
metal-semiconductor. An estimation of the NLH current in this injection
layer is the subject of present section. First of all, let us consider the
equilibrium electron density in the semiconductor side. Free energy of
electron gas per unit area of the surface has the following form
\cite[Liebsch]{6}
\begin{eqnarray}
&F =\int\limits_{0}^{\infty }f(\rho ,T)dz+2\pi e^{2}\int\limits_{0}^{\infty
}\rho (z)dz\int\limits_{0}^{\infty }z\rho (z)dz-  \nonumber \\
&e^{2}\pi \int\limits_{0}^{\infty }\int\limits_{0}^{\infty
}\rho (z)|z-z^{\prime }|\rho (z^{\prime })dzdz^{\prime }+\psi
_{0}\int\limits_{0}^{\infty }\rho (z)dz  \label{44}
\end{eqnarray}
where $f=(\rho (z),T)$ is a density of the free energy of injected electron
gas, which is a function of electron density $\rho (z)\ $and temperature $T$%
. At small density $\rho (z)\ $the $f=(\rho ,T)$ is a free energy of ideal
classical gas.

\[
f(\rho ,T)=T\rho \ln \rho /{\rho }_{T}
\]

We use the units where Boltzmann constant $k_{B}=1$. $\Psi _{0}$ is.a work
function for transfer electron from metal to the semiconductor. Minimization
of $F$\ with respect to $\rho (z)$ leads to the well known equation for
equilibrium density $\rho _{e}(z)$

\begin{eqnarray}
{\displaystyle{{\partial f} \over {\partial \rho }}}%
&+{2}\pi e^{2}z\int\limits_{0}^{\infty }\rho _{e}(z)dz
-e^{2}2\pi
\int\limits_{0}^{\infty }\rho _{e}(z^{\prime })|z-z^{\prime }|dz^{\prime }
\nonumber \\
&+2\pi e^{2}\int\limits_{0}^{\infty }z\rho _{e}(z)dz+\Psi _{0} =0\nonumber\\
\label{45}
\end{eqnarray}

For Boltzmann electron gas the solution of \ (45) looks as following

\begin{equation}
\rho _{e}(z)={\frac{z_{d}q_{\infty }^{2}}{(zz_{d}q_{\infty }+1)^{2}}}
\label{46}
\end{equation}
Here \ $q_{\infty }=\int_{0}^{\infty }\rho (z)dz\ $and

\[
z_{d}={%
{\displaystyle{2\pi e^{2} \over T}}%
},
\]
is full injected charge per unit of area of surface. An equality of chemical
potential of the metal and the semiconductor on the surface of the metal

\begin{equation}
\frac{{\partial f}}{{\partial \rho }}\Biggl|_{\rho =\rho _{e}(0)}=-\Psi _{0}
\label{47}
\end{equation}
gives
\begin{equation}
\rho _{e}(0)=\rho _{T}e^{-\Psi /T}  \label{48}
\end{equation}
and
\begin{equation}
q_{\infty }=(\rho _{T}/z_{d})^{1/2}e^{-\Psi /2T}  \label{49}
\end{equation}
In fact, the equations (45-49) summarise briefly the simplest theory of
Schottky barrier on the metal-semiconductor interface \cite{8}. Under
action of time-periodic external electric field $E_{0}\sin {\omega t}$ the
non-linear oscillations of the density $\rho (z,t)$ takes place. A system of
equations describing these oscillations looks very much like that of the
metal-vacuum interface equations (26-29), but the background positive charge
$\rho _{0}$ in Eq.(29) has to be taken zero.

Another important difference between the metal-insulator and
metal-semiconductor cases is connected with boundary conditions for
velocities $v(z,t)$ and $u(z,t)$ at $z=0$. There are two limit cases. In the
first one, the exchange of charges over the surface is quite slow which
means that total charge in the semiconductor is conserving under action of
the AC electric field and $v(0,t)=0$, $u(0,t)=0$ at the boundary. If we
neglect the viscosity a total NLH current vanishes.

Let us outline briefly the solution of the hydrodynamics equations (26-29)
in this case. At small electric $E_{0}$ and magnetic $H$\ fields and small
viscosity of electron liquid the first equation (26) can be rewritten as
\begin{equation}
\rho \omega _{\mu }=-c_{e}^{2}{\frac{{\partial \rho }}{{\partial z}}}+{\rho
4\pi (\rho (z)-\rho }_{{\infty }}{)+eE(t)\rho }  \label{50}
\end{equation}
where
\begin{equation}
{\rho (z,t)=}\int_{0}^{\infty }\rho (z)dz  \label{51}
\end{equation}
and
\begin{equation}
c_{e}^{2}={\rho \frac{{\partial }^{{2}}{f}}{{\partial \rho }^{{2}}}}
\label{52}
\end{equation}
For Boltzmann gas $c_{e}^{2}=\sqrt{T/m}$\ . The equation of continuity can
be rewritten as

\begin{equation}
{\frac{{\partial q}}{{\partial t}}=-v\rho }  \label{53}
\end{equation}
In the equation for $u(z,t)$\ we have to keep small terms connected with the
magnetic field (Lorentz force) and viscosity. Then, this equation can be
rewritten in the following form

\begin{equation}
{\frac{{\partial u\rho }}{{\partial t}}+\frac{{\partial }}{{\partial z}}%
vu\rho ={\omega }_{{L}}v\rho -{\omega }\mu u\rho +\frac{{\eta }}{{m}}\frac{{%
\partial }^{{2}}{u}}{{\partial z}^{{2}}}}  \label{54}
\end{equation}

Having in mind that we are interested in the full surface current averaged
over period of oscillation of external electric field

\begin{equation}
I_{H}{=e}\int_{0}^{\infty }\overline{\rho u}dz  \label{55}
\end{equation}
we arrive to the expression

\begin{equation}
I_{H}{=}\frac{{e}}{{\omega _{\mu }}}\int_{0}^{\infty }\frac{{\eta }}{{m}}{%
\frac{{\partial }^{{2}}{u}}{{\partial z}^{{2}}}}dz  \label{56}
\end{equation}
Here again the bar means the averaging over period of oscillation. We took
into account also that

\begin{equation}
\overline{{j}_{{z}}}=\overline{v\rho }{=0}  \label{57}
\end{equation}
due to the second equation.

As a rule, the viscosity of electronic gas $\eta $ depends on the density of
electron and, thus, depend on $z$\ . It is more convenient to rewrite (56) as

\begin{equation}
I_{H}{=-}\frac{{e}}{{\omega _{\mu }}\ m}\int_{0}^{\infty }({\frac{{\partial
\eta }}{{\partial \rho }}})\frac{{\partial \rho }}{{\partial z}}{\frac{{%
\partial u}}{{\partial z}}}dz  \label{58}
\end{equation}
Further we shall show that

\[
\frac{{\partial u}}{{\partial z}}\Biggl|_{z=0}=0
\]
At small viscosity and small ${\omega /\omega _{\mu }}$\ (low frequency of
external field)

\[
{u=\frac{{{\omega }_{{L}}}}{{\omega _{\mu }}}v}
\]
and we can express the $I_{H}$\ via normal velocity $v(z,t)$\ and density $%
\rho (z,t)$\ found from equation (50)

\begin{equation}
I_{H}{=-}\frac{{e{\omega }_{{L}}}}{m{\omega _{\mu }}^{{2}}}\int_{0}^{\infty
}({\frac{{\partial \eta }}{{\partial \rho }}})\frac{{\partial \rho }}{{%
\partial z}}{\frac{{\partial u}}{{\partial z}}}dz  \label{59}
\end{equation}
The viscosity of electron gas is unknown function of density of the gas.
Father, for estimation, we assume that is $%
{\displaystyle{{1} \over {m}}}%
{\displaystyle{{\partial \eta } \over {\partial \rho }}}%
$ is approximately equal to kinematic viscosity of electronic gas,

\[
\kappa {=}%
{\displaystyle{{1} \over {m}}}%
{\displaystyle{{\partial \eta } \over {\partial \rho }}}%
,
\]
taken at equilibrium surface density of electron gas, ${\rho }_{eq}(0).$Thus
(60) turns into the following expression

\begin{equation}
I_{H}{=-}\frac{{e{\omega }_{{L}}}}{{\omega _{\mu }}^{{2}}}\kappa
\int_{0}^{\infty }dz\frac{{\partial \rho }}{{\partial z}}{\frac{{\partial v}%
}{{\partial z}}}  \label{60}
\end{equation}
which we will use for estimation of nonlinear Hall surface current.

The solution of the eqs. (50) will be performed by theory of perturbation
with respect to a small amplitude of electric field $E_{0}$ . After
introduction of dimensionless time $\tau =\omega t$\ , coordinate $z={%
{\displaystyle{{c_{{e}}\xi } \over \sqrt{{\omega \omega _{\mu }}}}}%
}$, electric field ${\epsilon }_{{0}}={%
{\displaystyle{eE_{0} \over m{c_{{e}}}\sqrt{{\omega \omega _{\mu }}}}}%
}$\ and new function $\lambda (z,t)$.

\begin{equation}
q_{(z,t)}=q_{\infty }(1-\alpha \lambda (z,t))  \label{61}
\end{equation}
the (50) turns into the following equation

\begin{equation}
{\frac{{\partial }\lambda }{{\partial \tau }}=\frac{{\partial }^{{2}}\lambda
}{{\partial \xi }^{{2}}}+2\eta \frac{{\partial }\lambda }{{\partial \xi }}%
+\epsilon \frac{{\partial }\lambda }{{\partial \xi }}\cos \tau }  \label{62}
\end{equation}
which has to be solved with boundary conditions

\[
\lambda (\xi )\rightarrow 0\ \ \text{at}\ \ \xi \rightarrow \infty
\]

\[
\lambda (0)=\frac{{1}}{{\alpha }}
\]
Here we use the following notation

\[
{\alpha }={%
{\displaystyle{\sqrt{{\omega \omega _{\mu }}}{c_{{e}}}m_{0} \over 2\pi q_{\infty }e^{2}}}%
,}\beta ={%
{\displaystyle{{c_{{e}}} \over \sqrt{{\omega \omega _{\mu }}}}}%
}
\]

\[
{v}=-\beta \omega (\frac{{\partial }\lambda }{{\partial \tau }})/{(\frac{{%
\partial }\lambda }{{\partial \xi }})}
\]
Note that we are looking for the periodic solution of (63). If we make a
substitution

\begin{equation}
\lambda =-({\frac{{\partial \varphi }}{{\partial \xi }})/\varphi }
\label{63}
\end{equation}
the nonlinear eq. (63) turns into linear one

\begin{equation}
{\frac{{\partial \varphi }}{{\partial \tau }}=\frac{{\partial }^{{2}}{%
\varphi }}{{\partial \xi }^{{2}}}+\epsilon \cos \tau \frac{{\partial \varphi
}}{{\partial \xi }}}  \label{64}
\end{equation}
with the boundary conditions

\[
{\varphi }(\xi )\rightarrow \infty \ \ \text{at}\ \ \xi \rightarrow \infty
\]

\[
\frac{{\partial \varphi }}{{\partial \xi }}\Biggl|_{{\xi }=0}=\frac{{1}}{{%
\alpha }}{\varphi (0)}
\]

The perturbative analysis of linear Eq. (65) is rather trivial and we give
only the final result

\begin{equation}
\lambda (\xi ,t)=\lambda _{0}(\xi ){+\epsilon }\lambda _{1}(\xi ,t)
\label{65}
\end{equation}

\begin{equation}
\lambda _{0}=\frac{{1}}{\xi +{\alpha }}  \label{66}
\end{equation}
\begin{equation}
\lambda _{1}={\epsilon }\left[ \frac{{1}}{\xi +{\alpha }}\frac{{\partial
a(\xi )}}{{\partial \xi }}-\frac{{1}}{(\xi +{\alpha )}^{{2}}}{a(\xi )}\right]
e^{i\tau }  \label{67}
\end{equation}
where ${a(\xi )}$\ has the form

\begin{equation}
{a(\xi )}=i\left[ \frac{{1}}{1+k{\alpha }}e^{-k\xi }-1\right] ,k=\frac{{i+1}%
}{\sqrt{2}}  \label{68}
\end{equation}
Omitting intermediate calculations we give the expression for $I_{H}$\ in
the following form

\begin{equation}
I_{H}{=-e}\frac{b{{\omega }_{{L}}}}{{\omega _{\mu }}^{{2}}}\frac{q_{\infty
}\alpha }{\beta }{\epsilon }^{{2}}\int_{0}^{\infty }\frac{{1}}{(\xi +{\alpha
)}^{{4}}}v({\xi })d{\xi }  \label{69}
\end{equation}
where

\[
v({\xi })=\overline{\left( \frac{{\partial \eta }_{1}}{{\partial \tau }}{%
\frac{{\partial \eta }_{1}}{{\partial \xi }}}\right) }/\left( {\frac{{%
\partial \eta }_{0}}{{\partial \xi }}}\right) ^{2}
\]
since is proportional to $\sqrt{{\omega }}$\ ,the function $R(\alpha )$ ,

\begin{equation}
R(\alpha )=\int_{0}^{\infty }\frac{{1}}{{(\xi +\alpha )}^{{4}}}v({\xi })d{%
\xi }  \label{70}
\end{equation}
is actually a function of frequency of external electric field $\omega $\.
 Numerical calculation shows that $R(0)=-0.21$. Collecting all multipliers
we get the following expression for the $I_{H}$

\begin{equation}
I_{H}{=-\frac{3{H}}{2{\pi c}}R(0)\omega \kappa \Biggl(\frac{{\mu }E_{0}}{{%
c}_{{e}}}\Biggr)}^{{2}}  \label{71}
\end{equation}
As an example let us choose reasonable parameters

\[
{\rho }_{{\infty }}=10^{12}%
{\displaystyle{{1} \over {cm}^{{3}}}}%
,~~~~{\kappa }=10^{3}~{\frac{cm^{2}}{sec}},
\]
\[\mu =10^{5}~\frac{cm^{2}}{%
V\sec },~~~~{v}=10^{7}~{\frac{cm}{sec}}.
\]
Then, $I_{H}$\ can be written as

\[
I_{H}{=10}^{{-5}}{H\omega }E_{0}^{2}
\]
where $I_{H}$ is expressed in a $%
{\displaystyle{\mu A \over cm}}%
$ , $H$ in a $Oe$, $E_{0}$ in  a $V/cm$ and ${\omega }$\ in a $MHz$.

\section{Magnetophoresis of neutral particles on the surface.}

Consider now the neutral particle (atom, molecule or any nano-size particle)
located on the metal surface under joint action of normal AC-electric and
tangent permanent magnetic fields. As it has been shown in Section 1 the
driving force which causes the directed transport of particle on the surface
depends on a mass, charge and details of an interaction between particles
and the surface of metal. If negative and positive charges coupled in a
neutral pair as it takes place in atoms and molecules then the driving
forces acting on positive and negative charges are directed on opposite
sides, but their values, generally speaking, can be different. Thus, the
resulting force acting on the particle is not necessary zero leading to the
directed transport of the particles. As it is well-known the Lorentz forces
acting on the neutral particle in a magnetic field $H$ can be written as
\begin{equation}
{\bf F}_{l}=\frac{1}{c}\left[ \frac{{\partial }{\bf D}}{{\partial }t}\times
{\bf H}\right]  \label{72}
\end{equation}
where ${\bf D}$ is a dipole moment of the system of the charged particles.
On the other hand, the dipole moment ${\bf D}$ of the neutral particle in an
external electric field ${\bf E}_{0}\cos (\omega t)$ is an algebraic sum of
a permanent dipole moment of the particle ${\bf D}_{0}$ and the dipole
moment ${\bf D}_{1}(t)$ induced by an external time-periodic electric field $%
{\bf E}_{0}\cos (\omega t)$. Thus, we have
\begin{equation}
{\bf D}(t)={\bf D}_{0}+\alpha (\omega ){\bf E}_{0}\cos (\omega t)  \label{73}
\end{equation}
where $\alpha (\omega )$ is a frequency depending polarisibility of the
particle. As a rule, for neutral particle absorbed on a metal surface vector
${\bf D}_{0}$ is directed normally to the surface. The potential of
interaction between the dipole particle and metallic surface $V(z)$,
consists of two terms \cite{6}.

The Coulomb attraction potential
\begin{equation}
V_{id}(z)=-\frac{{\bf D}^{2}}{4z^{3}\epsilon }  \label{74}
\end{equation}
and a repulsive exchange potential $V_{ed}(z)$, which we again (see Section
1) choose in an exponential form
\begin{equation}
V_{ed}(z)=c_{d}a_{d}\exp \left( -\frac{z-z_{d}}{a_{d}}\right)  \label{75}
\end{equation}

Here $a_{d}$ is a radius of exchange forces on the metallic surface and $%
z_{d}$ is an equilibrium distance between the particle and the surface, $%
c_{d}$ is determined from an equilibrium condition
\begin{equation}
V(z)=V_{id}(z)+V_{ed}(z),\qquad \left. \frac{\partial V}{\partial z}\right|
_{z=z_{d}}=0  \label{76}
\end{equation}

As in the case of ion-metal system the friction coefficient, $\nu $, for
the dipole particle located on a metal surface depends on the distance $z$
(see \cite{6}). At large distances $z$, the $\nu (z)$ is proportional to $%
1/z^{3}$.

Similar to Section 1 we choose this dependence in a form
\begin{equation}
\nu (z)=\nu _{0}[1-\gamma _{d}(z-z_{d})]  \label{77}
\end{equation}
where $\nu _{0}=\nu (z_{d})$ and $\gamma _{d}=-\left.
{\displaystyle{\partial \nu (z) \over \partial z}}%
\right| _{z=z_{d}}$.

At small deviation from the equilibrium position of the particle the
equations of motion for its gravity center can be written as following

\begin{eqnarray}
M\ddot{z} &=&-M\nu (z)\dot{z}-\omega _{d}^{2}M(z-z_{d})-Q_{d}E_{0}(t)
\nonumber \\
M\ddot{x} &=&-M\nu (z)\dot{x}+\frac{H}{c}\frac{\partial D}{\partial t}
\label{78}
\end{eqnarray}

Here $\omega _{d}$ is an eigenfrequency of particle vibration on the surface
and
\[
Q_{d}=\frac{3D_{0}\alpha (\omega )}{z_{d}^{4}}
\]

Using (73) we get
\begin{equation}
\frac{\partial D}{\partial t}=-\omega \alpha (\omega )E_{0}\sin (\omega t)
\label{79}
\end{equation}

The system of equations (78) is complete analog of that in the Section
1. Omitting the details of the calculation we give the expression for
tangent velocity $\frac{dx}{dt}$ averaged over a period of oscillation of
the external effective electric field
\begin{equation}
\overline{\dot{x}}=v(\omega )=\frac{\gamma }{4}\left[ g_{0}(\omega
)g_{1}^{\ast }(\omega )+g_{1}(\omega )g_{0}^{\ast }(\omega )\right]
\label{80}
\end{equation}
where complex functions $g_{0}(\omega )$ and $g_{1}(\omega )$ have form
\begin{eqnarray}
g_{0}(\omega ) &=&\frac{Q_{d}E_{0}}{M(\omega _{d}^{2}-\omega ^{2}+i\nu
_{0}\omega )}  \label{81} \\
g_{1}(\omega ) &=&\frac{E_{0}\alpha (\omega )Hi\omega }{Mc(i\omega +\nu
_{0})}  \label{82}
\end{eqnarray}
Final expression of the $v(\omega )$ has a form
\begin{equation}
v(\omega )=\gamma \frac{3D_{0}}{z_{d}^{4}M^{2}c}\frac{\alpha ^{2}(\omega
)E_{0}^{2}H\omega ^{2}}{\omega ^{2}+\nu _{0}^{2}}\frac{\omega _{d}^{2}-\nu
_{0}^{2}-\omega ^{2}}{\left( \omega _{d}^{2}-\omega ^{2}\right) ^{2}+\nu
_{0}^{2}\omega ^{2}}  \label{83}
\end{equation}

At small $\omega $ the velocity $v(\omega )$ is decreasing as $\omega ^{2}$
in complete analogy with corresponding velocity of charged particle.
However, in difference with the Section 1 $v(\omega )$ changes its sign at $%
\omega =\sqrt{\omega _{d}^{2}+\nu _{0}^{2}}$. The general form of $v(\omega
) $ as a function of frequency is presented on Fig.4.

If the dipole moment of the particle $D_{0}$ is proportional to its volume
(as it takes place for ferroelectric medium) then $v(\omega )$ does not
change much with the increase of the radius of the particle and its mass.

If the external electric field $E_{0}(t)$ is a sum of two harmonics with
different frequencies $\omega _{1}$ and $\omega _{2}$ the resulting
velocity is additive,
\begin{equation}
v(\omega )=v_{1}(\omega _{1})+v_{1}(\omega _{2})  \label{84}
\end{equation}

For atomic size particles (atom or molecule) the maximal velocity could be
estimated as
\begin{equation}
v=10^{-12}E_{0}^{2}H  \label{85}
\end{equation}
where $v$ is given in cm/sec, $E_{0}$ in V/cm and $H$ in Oe.

\section{NLH effect phenomenology.}

In previous Sections we have shown that the joint action of the permanent
magnetic and AC-electric fields leads to the direct transport of particles
located nearly surface. An important role in consideration plays gradient of
the damping constant the system metal-particle. Generalizing all examples of
the previous Section we can assume that the NHL effect takes place in a
bulk of the material if there exists a non-zero gradient of conductivity
tensor
{$
j_{i}$ could be chosen as

\begin{equation}
j_{i}=\lambda (\omega ){\frac{{\partial \sigma _{ij}}}{{\partial z_{k}}}}{%
\dot{E}}_{k}{\dot{E}}_{l}\epsilon _{jlm}H_{m}  \label{86}
\end{equation}
where ${\dot{E}}_{k}$ is a time derivative of an external electric field and
$H$ is magnetic field. In a case of isotropic medium $\sigma _{ij}=\delta
_{ij}\sigma $ the density of current is
\begin{equation}
{\bf j}=\lambda ({\bf \nabla }\sigma {\dot{{\bf E}}})[{\dot{{\bf E}}}\times
{\bf H}]  \label{87}
\end{equation}
Here $\lambda $ is a phenomenological coefficient depending on frequency and
 material parameters.

To demonstrate the physical consequences of (88) consider the cylindrical
system shown on Fig.5. The time periodic voltage is applied between inner
and external surface of the cylindrical sample. The permanent magnetic
fields $H$\ is directed along main axes of the cylinder. We assume also that
the conductivity of the metal decreases along the radius of the sample, ${%
{\displaystyle{{\partial \sigma } \over {\partial r}}}%
<0}$.

In the chosen cylindrical geometry there exist only angular component of
the current density $j_{\varphi }(r)$\ which according to (88) can be
expressed as

\begin{equation}
j_{\varphi }(r)=\lambda (\omega ){\frac{{\partial \sigma }}{{\partial r}}%
\dot{{\bf E}}}^{{2}}{\bf H}  \label{88}
\end{equation}

Now, let us assume also, that the conductivity ${\sigma }$ , $%
\omega $\ and thickness of the sample are chosen such a way that the
electric field $E(t)$\ penetrates over the sample.Then the additional
magnetic field $H_{ad}(r)$\ induced by this current can be written as

\begin{equation}
H_{i}(r)=-{\frac{4\pi }{{c}}}\int_{r_{1}}^{r}\overline{j_{\varphi }}(r
){dr
= \frac{4\pi }{{c}}}(r-r_{1})\overline{j_{\varphi }}
\label{89}
\end{equation}
where $r_{1}$\ is an radius of the sample. There exist also diamagnetic
contribution to the induced magnetic field which is proportional to
diamagnetic susceptibility, $\chi $\ ,of the material

\begin{equation}
H_{d}=\chi H  \label{90}
\end{equation}

Summing up both terms we come to the conclusion that the full induced
magnetic field is positive if

\begin{equation}
-\lambda {\frac{{\partial \sigma }}{{\partial r}}}\overline{{\dot{{\bf E}}}^{%
{2}}}{\frac{4\pi }{{c}}}l_{d}>|\chi |  \label{91}
\end{equation}

Actually, the condition (91) is a condition of stability of the system to
the generation of magnetic field in the sample under action of time periodic
electric field $E(t)$.The details of phenomenon will be considered in the
future.

The main obstacle for observation of the NLH effect in AC electric field
seems to be connected with an absorption of ultrahigh frequency
electromagnetic field into material. The time-periodic components of Hall
current which is linear with respect to external electric field is greater
than a permanent component of the NLH current and may lead to heating up of
the sample. On the other hand, at low (or ultralow) temperature the
absorption and warming up could be made small enough for such observation.


\bigskip

\newpage
{\bf Captions to Figures} \vspace{0.15in}

Fig.~1. Frequency dependence of the velocity (20) of a single charged
particle at $\nu_0=0.1\o_e$ for three different masses $m=M/M_0=0.25$,
$M/M_0=1.0$ and $M/M_0=4.0$.

Fig.~2.  An attractor of the system (13-17). The A and A' are points of
return in x direction. Arrows on the trajectory show the direction of
the particle rotation, $\psi$ is an angle of inclination of the ellipse.

Fig.~3. Frequency dependence of $\g_1(\o )$ and $\g_2(\o )$  at
$\o_{\mu}=0.1\o_s$.

Fig.~4. Frequency dependence of the velocity (83) of neutral particle
at $\nu_0=0.1\o_d$.

Fig.~5. The principle set up for observation of NHL effect. Electric field
$E_0(t)$ is directed along the radius of the cylinder sample.

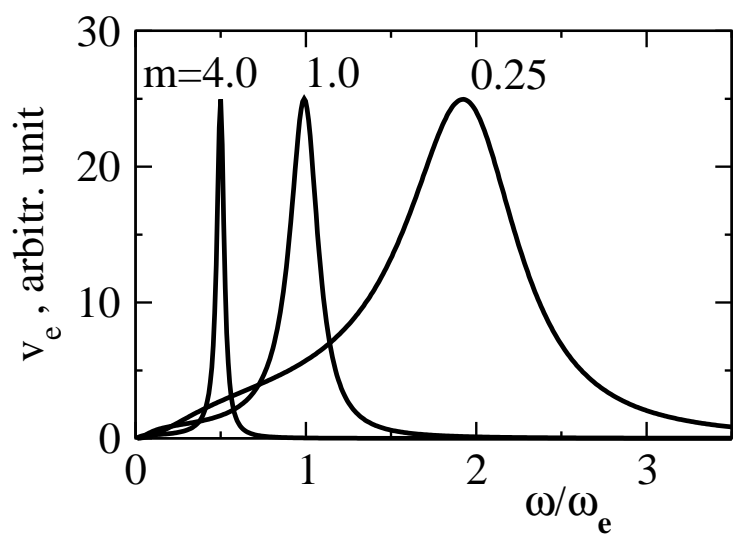
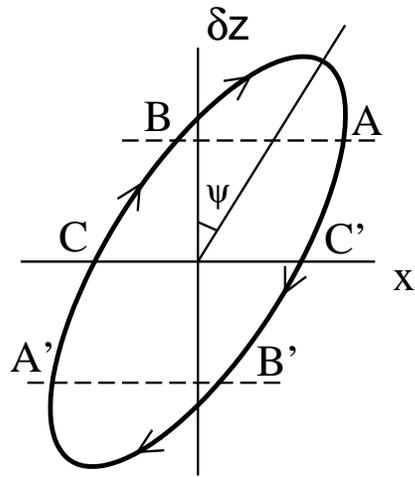
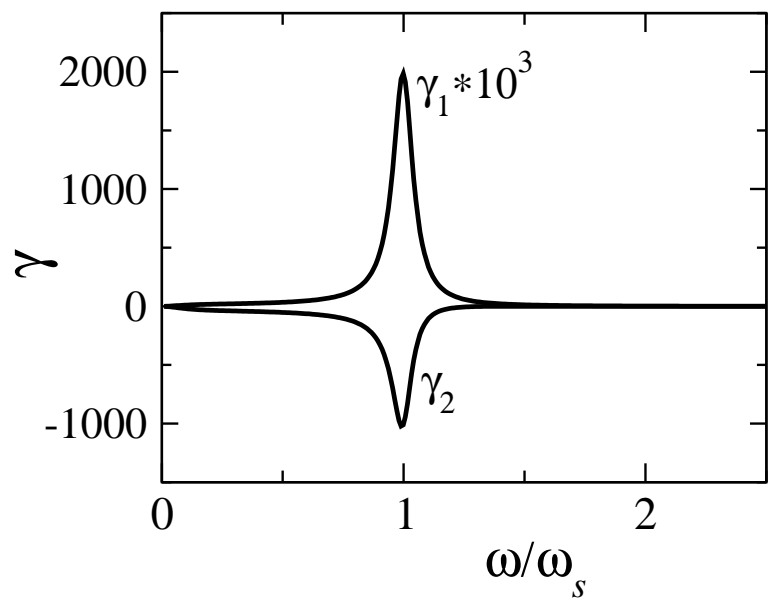
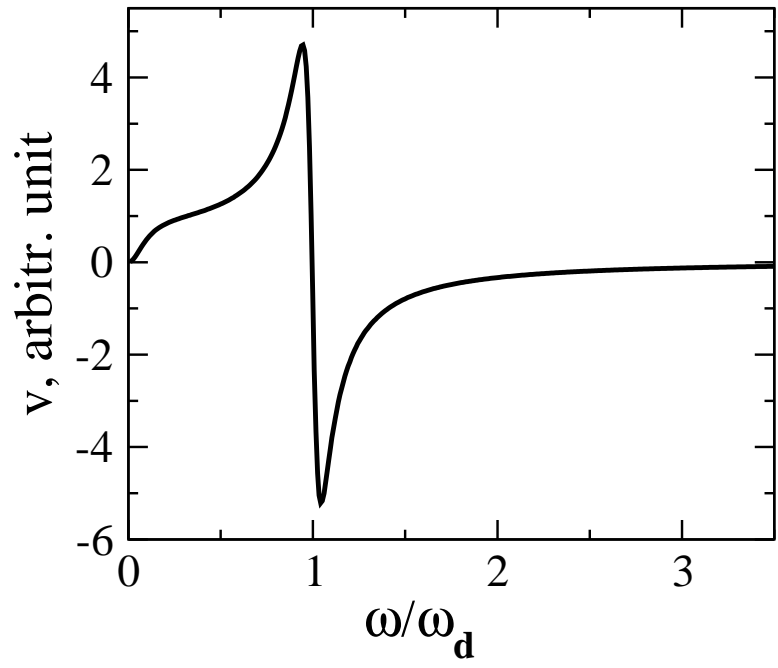
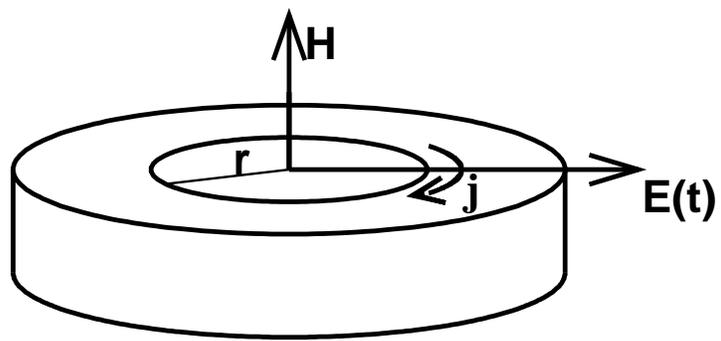
\end{document}